\documentclass[pra,showpacs,uplatex]{revtex4}

\usepackage{amsmath,amssymb}
\usepackage{bm}
\usepackage[dvipdfmx]{graphicx}
\usepackage[dvipdfmx]{color}
\usepackage[dvipdfmx]{epsfig}
\usepackage{float}
\usepackage{ulem}
\usepackage{color}

\begin{document}

\title{Random motion theory of an optical vortex in nonlinear birefringent media}

\author{Satoshi Tsuchida}
\email{tsuchida@gwv.hep.osaka-cu.ac.jp}
\address{Department of Physics, Osaka City University, Osaka, Osaka, 558--8585, Japan}

\author{Hiroshi Kuratsuji}
\address{Office of Professor Emeritus, Ritsumeikan University, Kusatsu, Shiga, 525-8577, Japan}

\date{\today}

\begin{abstract}

A theoretical study is presented for the random aspect of an optical vortex inherent in
the nonlinear birefringent Kerr effect, which is called the optical spin vortex.
We start with the two-component nonlinear Schr\"{o}dinger
equation. The vortex is inherent in the spin texture caused by
an anisotropy of the dielectric tensor, for which the role of spin
is played by the Stokes vector (or pseudospin).
The evolutional equation is derived for the vortex center coordinate
using the effective Lagrangian of the pseudospin field.
This is converted to the Langevin equation in the presence of the fluctuation together with the dissipation.
The corresponding Fokker-Planck equation is derived and analytically solved for a particular form of the
birefringence inspired from the Faraday effect.
The main consequence is that the relaxation distance for the distribution function
is expressed by the universal constant in the Faraday effect and the size of optical vortex.
The result would provide a possible clue for future experimental study in polarization optics from a stochastic aspect.

\end{abstract}

\pacs{42.25.Lc, 42.65.-k, 05.10.Gg}

\maketitle

\section{Introduction}

The vortex is a typical topological defect in condensed matter physics~\cite{LL}.
In optics, the vortex has been explored since the early development of nonlinear optics
\cite{Chiao, Swartz, Snyder, Christou, Rozas, LL2}.
The essence of the optical vortex relies on the Kerr effect,
which is described by the nonlinear Schr{\"o}dinger equation
for the single-component complex wave function.

On the other hand, it is well established that the polarization degree of freedom drastically changes
the feature of the optical phenomena \cite{LL2,BW}.
Indeed, wide viewpoints may be open for the physics of optical vortex.
Along this train of thought, by taking into account the polarization degree of freedom,
the more intricate form of a vortex can be expected, which is caused by nonlinear birefringence.
The original idea of the vortex inspired by polarization was suggested early on in the article~\cite{Kakigi}.
This specific type of vortex can be described by the ``Stokes parameters,"
which characterize the polarization degree of freedom.
Following an analogy between the Stokes parameters and a conventional spin,
it is natural to describe the new optical vortex in analogy with a spin vortex
condensed matter physics~\cite{Belavin,Kura1,Moon}.
In this sense, the vortex inspired from the nonlinear birefringence may be called
an ``optical spin vortex (OSV)."

A particular interest is to investigate the interaction
of vortex with the other degrees of freedom inherent in the medium.
Among the various interactions, it would be specifically intriguing to scrutinize
the random effect on an optical spin vortex.
To explore this problem is the purpose of the present article.

To begin with this attempt, we recall the general background:
The wave propagation in media generally suffers from the randomness
caused by the presence of various fluctuations built in the media and/or external agents
\cite{Dashen,Shapiro,Ishimaru,Segev,Someda,Biele,TK1}.
The topics cover a wide class of problems including quantum mechanics.
In a previous paper \cite{TK2}, we have investigated
a stochastic theory for the polarized light, in which we presented the
Langevin and Fokker-Planck equations for the {\it integrated form} for the Stokes parameters,
that is inspired by the single-component scalar vortex \cite{Akhmediev}.
Although the present problem is quite different from the dynamics of the Stokes parameters itself,
it shares the same spirit:
Namely, our aim is to investigate the random theory OSV in the framework of ``fluctuation" and ``dissipation."
The essence is as follows:
The field of the Stokes parameter and its angular representation play a role of paramount importance, which
just validates the stochastic (random) theoretic description of the optical spin vortex.
On the basis of this viewpoint,
the evolution equation of motion for the vortex center can be converted the Langevin equation,
which is described by the Brownian motion of the vortex.
This yields the Fokker-Planck equation.
The present attempt is expected to shed light on the hidden aspect
in the discipline of the optical vortex from the stochastic viewpoint.

This paper is organized as follows:
In the next section we present an overview of the OSV.
The basic equation is given by the two-component nonlinear Schr{\"o}dinger equation (abbreviated as TCNLS)
which is derived from the Maxwell equation by adopting an envelope approximation.
The crux is that the evolution is governed by the propagation distance which plays a role of time.
In Sec. \ref{sec:evolution} the evolutional equation for the vortex center is given
on the basis of the effective Lagrangian for the Stokes parameters.
The procedure given there is the modified of the previous report \cite{Kakigi}.
In Sec. \ref{sec:random} the Langevin equation
for the OSV, and this is converted to the Fokker-Planck equation.
In Sec. \ref{sec:analysis} an analysis will be presented in an analytical manner.

\section{Preliminary: Effective Lagrangian for the pseudospin field}
\label{sec:preliminary}

Following the previous works \cite{Kura2,Kakigi},
we start with a brief sketch for the basic formulation of the
polarization optics, which is based on the ``para-axial scheme."
Let us consider the two-component wave function for the polarized light,
which is written in terms of the circular polarized basis, $ \psi = {}^t\left(\psi_1,\, \psi_2\right) $.
The wave function $ \psi $ satisfies the Schr{\"o}dinger-type equation:
\begin{eqnarray}
  \label{eq:sch1}
  i \lambda \frac{ \partial \psi }{ \partial z}
  =
  \left( - \frac{ {\lambda}^2 }{ 2 n_{0} } {\nabla}^2 + \hat{V} \right) \psi
  \equiv
  \hat{H} \psi,
\end{eqnarray}
where $ \lambda $ is the wavelength divided by $ 2 \pi $,
$ n_{0} $ means the refractive index for the isotropic media, and $ \hat{H} $ is the Hamiltonian.
We note that $ z $ plays a role of time in Eq.~(\ref{eq:sch1}),
and $ \hat{V} $ represents the ``field-dependent'' potential.
In the following argument we adopt the contribution from the nonlinear Kerr birefringence~\cite{Maker}:
\begin{eqnarray}
  \hat{V}
  =
  - g \left(
  \begin{array}{cc}
    - \frac{ \vert \psi_1 \vert^2 - \vert \psi_2 \vert^{2} }{ 2 }  &  {\psi}_{2}^{\dagger} \psi_1 \\
     \psi_1^{\dagger} \psi_2  &  \frac{ \vert \psi_1 \vert^2 - \vert \psi_2 \vert^2 }{ 2 }
  \end{array}
  \right).
  \label{x}
\end{eqnarray}

We now introduce the ``quantum'' action leading to the
Schr{\"o}dinger-type equation, which is given by
\begin{eqnarray}
  I
  =
  \int {\psi}^{\dagger} \left( i {\lambda} \frac{ \partial }{ \partial z } - \hat H \right) \psi \, d^2 x dz
  \equiv
  \int L \, dz.
\end{eqnarray}
Having introduced the {\it Lagrangian}
$ L $ for the two-component field $\psi$, we rewrite this in terms
of the Stokes parameters, which is defined as
$ S_{k} = \psi^{\dagger} \sigma_{k} \psi $, $ S_0 = \psi^{\dagger} {\boldsymbol{1}} \psi $
with $ k = x, y, z $~\cite{Kakigi1,Brosseau}.
We note that the relation $ S_{0}^{2} = S_{x}^{2} + S_{y}^{2} + S_{z}^{2} $
holds, namely, $ S_{0} $ gives the field strength.
Using the spinor representation,
\begin{eqnarray}
  \psi_{1} = \sqrt{ S_{0} } \cos{ \frac{ \theta }{ 2 } } ~,~~
  \psi_{2} = \sqrt{ S_{0} } \sin{ \frac{ \theta }{ 2 } } \exp \left[ i \phi \right],
\end{eqnarray}
we have the polar form for the Stokes vector
\begin{eqnarray}
  {\boldsymbol{S}}
  =
  ( S_{x} , S_{y} , S_{z} )
  \equiv
  ( S_{0} \sin \theta \cos \phi , S_{0} \sin \theta \sin \phi , S_{0} \cos \theta ),
\end{eqnarray}
which forms a pseudospin and is given by the point on the Poincar\'{e} sphere.
In terms of the angle variables, the first term of the Lagrangian,
which is denoted as $ L_{C} $, is written as
\begin{eqnarray}
  L_{C}
  =
  - \frac{ S_{0} {\lambda} }{ 2 } \int \left( 1 - \cos \theta \right) \frac{ \partial \phi }{ \partial z } \, d^2 x,
\end{eqnarray}
which is called the {\it canonical term},
while the Hamiltonian becomes the sum of the kinetic and the potential energies, $ H = H_t + V $:
\begin{eqnarray}
  H_{t}
  &=&
  \int \frac{ S_{0} {\lambda}^{2} }{ 2 n_{0} }
  \left\{ \frac{1}{4} ( {\boldsymbol{\nabla}} \theta )^{2} + {\sin}^{2} \frac{ \theta }{ 2 } ( {\boldsymbol{\nabla}} \phi )^{2} \right\} d^2x, \\
  V
  &=&
  \int {\psi}^{\dagger} \hat V \psi d^2x
  =
  - \frac{g}{2} \int \left( S_{x}^{2} + S_{y}^{2} - S_{z}^{2} \right) d^2x.
\end{eqnarray}
Here we discard the term including the derivative of $ S_{0} $,
which means that the argument is restricted to the case that $ S_{0} $ becomes constant.
The physical meaning of the kinetic energy may be clarified by rewriting as
the sum of two terms: $ H_{t} = H_{1} + H_{2} $,
\begin{eqnarray}
  H_{1}
  &=&
  \int \frac{ S_{0} {\lambda}^{2} }{ 8 n_{0} } \left\{ ( 1 - \cos \theta ) ( {\boldsymbol{\nabla}} \phi ) \right\}^{2} d^2x, \nonumber \\
  H_{2}
  &=&
  \int \frac{ S_{0} {\lambda}^{2} }{ 8 n_{0} } \left\{ ( {\boldsymbol{\nabla}} \theta )^{2} + {\sin}^{2} \theta ( {\boldsymbol{\nabla}} \phi )^{2} \right\} d^2x .
\end{eqnarray}
Namely, defining the ``velocity field''
$ \boldsymbol{v} = ( 1 - \cos \theta ) {\boldsymbol{\nabla}} \phi $,
the first term is regarded as
fluid kinetic energy inherent in spin structure,
while the second term represents an intrinsic energy for the pseudospin,
which exactly coincides with a continuous Heisenberg spin chain \cite{Ono}.
In other words, if using the terminology with the elastic theory,
$ H_1 $ corresponds to the {\it twisting energy}, and $ H_2 $ to the {\it bending energy} \cite{elastic}.

\section{Vortex and its evolution}
\label{sec:evolution}

\subsection{Profile of polarization vortex}

We now construct an explicit form of a vortex solution for the TCNLS.
The solution we want here is a ``static'' solution, namely,
we look for the solution that is independent of the variable $ z $.

A possible candidate for the single vortex may be prescribed as follows.
As a phase function, we choose $ \phi = n \, {\tan}^{-1} \left( \frac{ y }{ x } \right) $,
with $ n = 1,2, \cdots $, being the winding number,
whereas the profile function $ \theta $ is given as a function of the radial variable
$ r ~( = \sqrt{ x^{2} + y^{2} } ) $.
Note that such a vortex becomes nonsingular,
namely, the velocity field does not bear the singularity due to the
behavior of $ \theta(r) $ near the origin (see below).
The static Hamiltonian is thus written in terms of the field $ \theta(r) $:
\begin{eqnarray}
  H
  =
  \frac{ S_{0} {\lambda}^2 }{ 8 n_{0} }
  \int \left[ \left\{ \left( \frac{ d \theta }{ dr } \right)^{2}
  + \frac{ 4 n^{2} }{ r^{2} } {\sin}^{2} \frac{ \theta }{ 2 } \right\} + g' {\cos}^{2} \theta \right] rdr.
\end{eqnarray}
Here we put $ g'= { 8 g n_0 S_0 \over \lambda^2 } $ as a positive value
and discard the constant term.
The profile function $\theta(r)$ may be derived
from the extremum of $ H $; namely, the Euler-Lagrange equation
leads to
\begin{eqnarray}
  \frac{ d^2 \theta }{ d {\xi}^2 } + \frac{ 1 }{ \xi } \frac{ d \theta }{ d \xi }
  - \frac{ n^{2} }{ {\xi}^2 } \sin \theta + \frac{ 1 }{ 2 } \sin { 2 \theta } = 0,
\end{eqnarray}
where we adopt the scaling of the variable: $ \xi = \sqrt{ g' } r $.
In order to examine the behavior of $ \theta ( \xi ) $,
we need a specific boundary condition at $ \xi = 0 $ and $ \xi = \infty $.
We first consider the behavior near the origin $ \xi = 0 $,
for which the differential equation behaves like the Bessel equation,
so we see $ \theta(\xi) \simeq J_{ n } ( \xi ) $, which satisfies $ \theta (0) \simeq 0 $.
This means that the optical state is circular polarization at the origin.

We examine the behavior at $ \xi = \infty $.
This is simply performed by checking the stationary feature of the solution.
If putting $ \theta(\xi) = \frac{ \pi }{ 2 } + \alpha $, with $ \alpha $ the infinitesimal deviation,
then we have the linearized equation  $ {\alpha}'' - {\alpha} \simeq 0 $ near $ \xi = \infty $,
which results in $ \alpha \simeq \exp [ - \xi ] $.
This means that the solution should approach $ \theta ( \infty ) = \frac{ \pi }{2} $.
Thus, the state becomes linear polarization at infinity.

From above consideration, we have a profile for the optical vortex, such that
the interpolation
\begin{eqnarray}
  \theta(\xi) = \frac{ \pi }{ 2 } \left\{ 1 - f(\xi) \right\},
\end{eqnarray}
where function $ f(\xi) $ satisfies $ f(0) = 1 $ and $ f(\infty) = 0 $.
Then, for a typical example of $ {\theta} (\xi) $,
we can choose the form of $ \cos \theta (\xi) $ to be a Gaussian function.
Here it should be noted that the rotational behavior of the vortex can be governed
by angle $ \phi \propto {\tan}^{-1} \left( \frac{y}{x} \right) $.
From this expression, we see that the origin ($ r = 0 $) possesses a sort of singularity.

\subsection{Evolution equation of the vortex}

Having constructed the explicit form for the
vortex solution, we now consider the evolutional behavior
for a single vortex with respect to the propagation direction $ z $.
Following the procedure used in the magnetic vortex~\cite{Ono},
let us introduce the coordinate of the center of vortex,
$ {\boldsymbol{R}} (z) = ( X(z), Y(z) ) $,
by which the vortex solution is parameterized such that
$ \theta \left[ {\boldsymbol{x}} - {\boldsymbol{R}} (z) \right] $ and
$ \phi \left[ {\boldsymbol{x}} - {\boldsymbol{R}} (z) \right] $.
By using this parametrization and noting the chain rule,
$ \frac{ \partial \phi }{ \partial z } = \frac{ \partial \phi }{ \partial {\boldsymbol{R}} } \dot{\boldsymbol{R}} $,
$ \frac{ \partial \phi }{ \partial {\boldsymbol{R}} } = - {\boldsymbol{\nabla}} \phi $
with $ \dot{\boldsymbol{R}} \equiv \frac{ d {\boldsymbol{R}} }{ dz } $,
the Lagrangian is reduced to the form written in terms of the vortex center:
\begin{eqnarray}
  {\tilde{L}}
  =
  {\tilde{L}}_{C} - {\tilde{H}}
  =
  \frac{ S_{0} \lambda }{ 2 } \int {\boldsymbol{v}} \cdot \dot{ \boldsymbol{R} } \, d^2 x - {\tilde{H}}.
\end{eqnarray}
Here $ {\tilde{L}}_C $ corresponds to the canonical term,
while $ \tilde{H} $ is a reduced Hamiltonian, which will be given below.
Using the Euler-Lagrange equation, we obtain the ``equation of motion" as follows:
\begin{eqnarray}
  \label{nine}
  \frac{ S_0 \lambda }{ 2 } \sigma
  \left( \boldsymbol{k} \times \dot{\boldsymbol{R}}\right)
  =
  \frac{ \partial \tilde{H} }{ \partial {\boldsymbol{R}} },
  \label{gyration1}
\end{eqnarray}
which will be derived in the Appendix A.
Here $ {\boldsymbol{k}} $ is the unit vector perpendicular
to the $xy$ plane and $ \sigma $ means the topological index characterizing the OSV.
We also have a partner equation:
\begin{eqnarray}
  - \frac{ S_0 \lambda }{ 2 } \sigma \dot{\boldsymbol{R}}
  =
  \boldsymbol{k} \times \frac{ \partial \tilde{H} }{ \partial {\boldsymbol{R}} }.
 \label{gyration2}
\end{eqnarray}
The right-hand side in Eq.~(\ref{gyration2}) represents a gyration,
which may be regarded as an analogy with the ``Lorentz force."

Next, we discuss the concrete form of the Hamiltonian $ \tilde{H} $.
The simple case may be that of an axial-symmetric form, namely,
$ {\tilde{H}} $ depends on $ \vert {\boldsymbol{R}} \vert $ only.
In what follows we restrict the argument to this case. Thus we write
\begin{eqnarray}
  \tilde{H} = \int {\psi}^{\dagger} \hat{v} {\psi} \, d^2x.
\end{eqnarray}
As the most tractable one, we consider the Hamiltonian that arises from the Faraday effect,
namely, $ \hat v = {\sigma}_z h_z({\boldsymbol{x}}) $,
for which $ \tilde{H} $ is written as following expression~\cite{Tsudagawa}:
\begin{eqnarray}
  \tilde{H}_{\rm{F}}
  =
  \int \gamma h_{z} ({\boldsymbol{x}}) S_{z} \left[ {\boldsymbol{x}} - {\boldsymbol{R}} (z) \right] \, d {\boldsymbol{x}},
\end{eqnarray}
where $ \gamma $ corresponds to Verde constant,
$ h_{z} $ means magnetic field of the $ z $-axis component,
and the index $ {\rm{F}} $ stands for the ``Faraday" effect.
We assume that $ h_{z} $ has a parabolic form,
which is used for confinement for the atomic gas in the Bose-Einstein condensation~\cite{BEC}.
In addition, it is plausible that the vortex profile,
as noted above, $ \cos \theta (= S_z) $ may be approximated to the Gaussian form:
\begin{eqnarray}
  \label{eq:hf}
  \tilde{H}_{\rm{F}}
  &\simeq&
  \int_{ - \infty }^{ \infty } \gamma {\boldsymbol{x}}^{2}
  \exp \left\{ - a \left[ {\boldsymbol{x}} - {\boldsymbol{R}} (z) \right]^{2} \right\} d {\boldsymbol{x}} \nonumber \\
  &=&
  k R^{2} + ( {\rm{ ~independent ~of } } ~R ~).
\end{eqnarray}
where $ k = \frac{ {\gamma} {\pi} }{ a } $, and $ a $ represents a size parameter for the vortex.
This form plays a crucial role in the analysis of the Fokker-Planck equation,
which will be given later.

As another candidate we mention the Hamiltonian due to {\it pinning} that is caused by the magnetic impurity.
Let us suppose that the impurity is placed at the origin,
which induces the interaction $ \hat v = c \hat \sigma_z  \delta({\boldsymbol{x}}) $,
and hence the Hamiltonian reads $ \tilde H_{\rm{pin}} (= c S_z({\boldsymbol{R}}) ) = c \exp[ -a {\boldsymbol{R}}^2] $,
which is axial symmetric.

\section{Random Effect}
\label{sec:random}

\subsection{Langevin equation for the vortex center}

Now we come to our main topics for the random effects on the motion of an optical spin vortex.
We give a brief explanation for the background of this context.

First to be mentioned is the general effect of propagation of light through a disordered medium.
If a vortex is created in the optical field, it is inevitably affected by this disorder.
As for the other causes, the vortex may be disturbed by the random
impurities giving rise to a linear birefringence and/or an optical activity.
During the scattering event of the light, the vortex is acted by uncontrollable birefringent effect.
By repeating this process, the vortex motion turns out to be random,
and this process is nothing but the Brownian motion of the vortex.
Furthermore one needs to take into account the {\it temperature fluctuation}
that is inherent in the medium under consideration.

In addition to the random fluctuations, it is inevitable that one takes account of the dissipative effect on the
vortex, which is caused by the energy absorption during the wave propagation,
namely, there is an energy loss during the vortex evolution.
In the present context, it is enough to adopt the phenomenological way as given below~\cite{Landau}.

Here a remark is in order: The disorder (fluctuation) should control the system internally by
balancing the fluctuation with the dissipation to maintain the distribution of the polarization in a proper way,
which is guaranteed by the well-known {\it fluctuation-dissipation} theorem.

Now we denote the random fluctuation as $ {\boldsymbol{\xi}} $ and assume the Gaussian noise; that is,
the relation $ \langle \xi_i(z+u)\xi_j(z) \rangle  = \kappa\delta_{ij}\delta(u) $ holds.
The dissipative effect may be included in a phenomenological form, $ \mu \dot {\boldsymbol{R}} $.
This dissipation is also required by the so-called fluctuation-dissipation theorem.

The ``fluctuation" and ``dissipation" can be incorporated in the evolution equations
(\ref{gyration1}) and (\ref{gyration2}), by replacing
$ \nabla \tilde H \rightarrow \nabla \tilde H + \mu\dot {\boldsymbol{R}}  + {\boldsymbol{\xi}} $,
that is,
\begin{eqnarray}
  \label{eq:el1}
  - \frac{ S_{0} \lambda }{ 2 } \sigma \dot{ {\boldsymbol{R}} }
  &=&
  {\boldsymbol{k}} \times \left( \frac{ {\partial} \tilde{H} }{ {\partial} {\boldsymbol{R}} }
  + {\mu} \dot{ {\boldsymbol{R}} } + {\boldsymbol{\xi}} \right), \\
  \label{eq:el2}
  \frac{ S_{0} \lambda }{ 2 }\sigma \left( {\boldsymbol{k}} \times \dot{\boldsymbol{R}} \right)
  &=&
  \left( {\partial \tilde H \over \partial \boldsymbol{R}}
  + {\mu} \dot{ {\boldsymbol{R}} } + {\boldsymbol{\xi}} \right).
\end{eqnarray}
From Eqs.~(\ref{eq:el1}) and (\ref{eq:el2}), we can derive an equation for $ \dot{ {\boldsymbol{R}} } $ as follows:
\begin{eqnarray}
  \label{eq:rdot}
  \dot{ {\boldsymbol{R}} }
  =
  - {\boldsymbol{A}} + {\boldsymbol{\eta}},
\end{eqnarray}
where  $ {\boldsymbol{A}} $  and $ {\boldsymbol{\eta}} $  are given by
\begin{eqnarray}
  {\boldsymbol{A}}
  &=&
  \frac{ 1 }{ \left( \frac{ S_{0} \lambda }{ 2 } \sigma \right)^{2} + {\mu}^{2} }
  \left[
  {\mu} \frac{ {\partial} \tilde{H} }{ {\partial} {\boldsymbol{R}} }
  +
  \left( \frac{ S_{0} \lambda }{ 2 } \sigma \right) \left( {\boldsymbol{k}} \times \frac{ {\partial} \tilde{H} }{ {\partial} {\boldsymbol{R}} } \right) \right], \\
  {\boldsymbol{\eta}}
  &=&
  - \frac{ 1 }{ \left( \frac{ S_{0} \lambda }{ 2 } \sigma \right)^{2} + {\mu}^{2} }
  \left[
  {\mu} {\boldsymbol{\xi}}
  + \left( \frac{ S_{0} \lambda }{ 2 } \sigma \right) \left( {\boldsymbol{k}} \times {\boldsymbol{\xi}} \right)
  \right].
  \label{eta}
\end{eqnarray}
This expression of $ {\boldsymbol{A}} $ is just the modified of the gyration term in (\ref{gyration2}),
namely, we have an extra term: the first term, which may be called the gradient term.
We have the relation,
$ \frac{ d \tilde{H}}{ dz } = \frac{ {\partial} \tilde{H} }{ {\partial} {\boldsymbol{R}} } \cdot \dot{ {\boldsymbol{R}} }
\leq 0 $, which means that the energy dissipates in association with the light propagation.

\begin{figure}[t]
  \begin{center}
    \vspace{20mm}
    \hspace{-45mm}
    \includegraphics[width=\columnwidth]{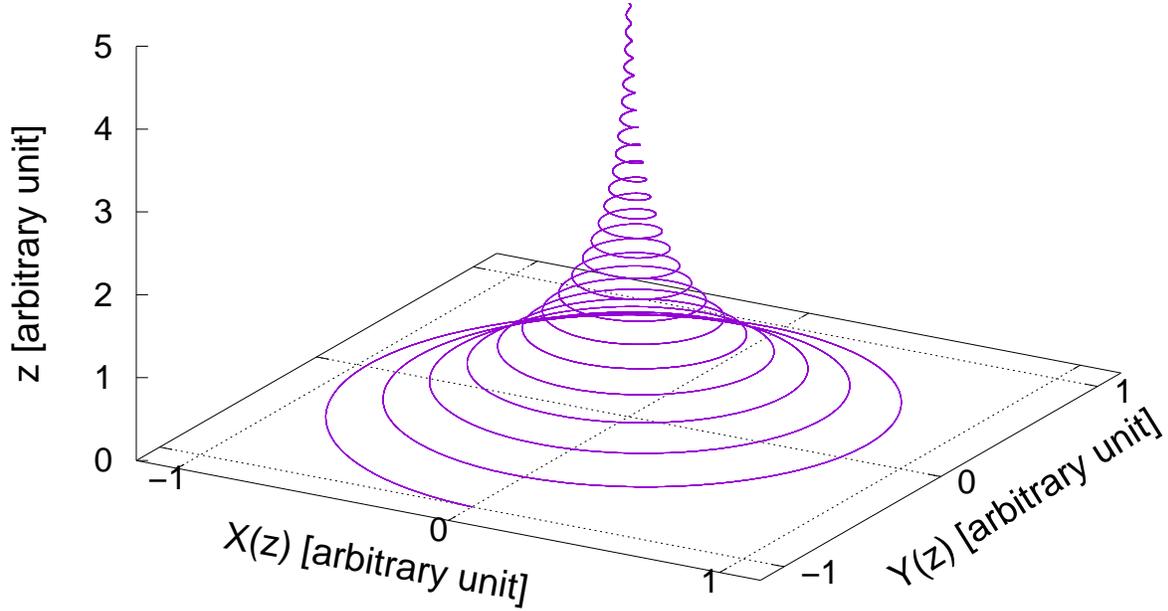}
    \vspace{-20mm}
    \caption{Schematic image of an ``inspiral."}
    \label{fig:inspiral}
  \end{center}
\end{figure}

{\it The limiting case without fluctuation}:
If we assume the limit of $ {\boldsymbol{{\eta}}} \rightarrow 0 $,
Eq.~(\ref{eq:rdot}) becomes as follows:
\begin{eqnarray}
  \frac{ dR }{ d {\Theta} } = \frac{ 2 \mu }{ S_{0} {\lambda} {\sigma} } R .
\end{eqnarray}
Especially, for the case of the Faraday effect,
the ``orbit" of the vortex center is given by following expressions:
\begin{eqnarray}
  \label{eq:sol_rdot}
  R(z)
  &=&
  R_{0} \exp \left[ - \frac{ 2 {\mu} k }{ \left( \frac{ S_{0} \lambda }{ 2 } \sigma \right)^{2} + {\mu}^{2} } z \right],  \nonumber \\
  \Theta(z)
  &=&
  - \frac{ 2 {\mu} k \left( \frac{ S_{0} \lambda }{ 2 \mu } \sigma \right) }{ \left( \frac{ S_{0} \lambda }{ 2 } \sigma \right)^{2} + {\mu}^{2} } z + {\Theta}_{0},
\end{eqnarray}
where $ R_{0} $ and $ {\Theta}_{0} $ are initial values of $ R $ and $ \Theta $, respectively.
Equation (\ref{eq:sol_rdot}) indicates that
the motion of the vortex center behaves as an ``inspiral,"
which corresponds to a spiral with decreasing a radius as is shown in Fig.~\ref{fig:inspiral}.
This implies that one can extract the information of the dissipation constant by
observing the decay rate of the radius.

\subsection{Fokker-Planck Equation}

The Langevin equation can be converted to the Fokker-Planck (FP) equation by utilizing
the ansatz of the Gaussian white noise for the reduced fluctuation.
That is, $ {\boldsymbol{\eta}} $ obeys the following relation~\cite{foot1}:
\begin{eqnarray}
  \langle {\eta}_{i} (z) {\eta}_{j} (z+u) \rangle
  =
  \frac{ \kappa }{ 1 + {\mu}^{2} } {\delta}_{ij} {\delta} (u)
  =
  2h {\delta}_{ij} {\delta} (u) .
\end{eqnarray}
The white noise is realized by the probability distribution of $ {\boldsymbol{\eta}} $
given by the standard Gaussian functional form:
\begin{eqnarray}
  \label{eq:GW}
  P [ {\boldsymbol{\eta}} (z) ] = \exp \left[ - \frac{1}{2h} \int_{0}^{z} {\boldsymbol{\eta}}^{2} (z) \, dz \right] .
\end{eqnarray}
In order to derive the FP equation, we employ the functional integral technique.
Following the procedure in Refs.~\cite{TK1, TK2},
we can obtain the final form of the FP equation written in the probability
distribution function $ P $ as follows
(for the details, see Appendix B):
\begin{eqnarray}
  \label{eq:FPeq}
  \frac{ {\partial} P }{ {\partial} z }
  =
  \frac{h}{2} \left( \frac{ {\partial} }{ {\partial} {\boldsymbol{R}} } \right)^{2} P
  +
  \frac{ {\partial} }{ {\partial} {\boldsymbol{R}} } \cdot
  \left( {\boldsymbol{A}} P \right).
\end{eqnarray}
The FP equation naturally gives the equation for the average value of a function $ F({\boldsymbol{R}}) $,
$ \langle F({\boldsymbol{R}}) \rangle = \int F({\boldsymbol{R}}) P({\boldsymbol{R}}, z) d {\boldsymbol{R}} $:
\begin{eqnarray}
  \frac{d \langle F \rangle}{dz}
  =
  \frac{h}{2} \langle {\nabla}^2 F \rangle - \langle {\boldsymbol{\nabla}} F \cdot {\boldsymbol{A}} \rangle.
\end{eqnarray}
As a special case, if we choose $ F = X $ or $ Y $, we recover the equation of motion for $ {\boldsymbol{R}} $.

Alternatively, the FP equation~(\ref{eq:FPeq}) can be expressed in a form of the continuity equation:
\begin{eqnarray}
  \frac{ {\partial} P }{ {\partial} z } + {\boldsymbol{\nabla}} \cdot {\boldsymbol{J}} = 0
\end{eqnarray}
where $ {\boldsymbol{J}} $ corresponds to a current, which is written in
terms of the polar coordinate basis $ ({\boldsymbol {e}}_R, {\boldsymbol {e}}_{\Theta} $).
That is, $ {\boldsymbol{J}} $ becomes
\begin{eqnarray}
  {\boldsymbol{J}}
  &=&
  - \left( \frac{h}{2} {\boldsymbol{\nabla}} P + {\boldsymbol{A}} P  \right)
  =
  J_{R} {\boldsymbol{e}}_{R} + J_{\Theta} {\boldsymbol{e}}_{\Theta}
\end{eqnarray}
with each component:
\begin{eqnarray}
J_{R}
&=&
- \frac{h}{2} \frac{ {\partial} P }{ {\partial} R }
- \frac{ P }{ \left( \frac{ S_{0} {\lambda} }{ 2 } {\sigma} \right)^{2} + {\mu}^{2} }
  \left( {\mu} \frac{ {\partial} {\tilde{H}} }{ {\partial} R }
  - \frac{ S_{0} {\lambda} }{ 2 } {\sigma} \frac{1}{R} \frac{ {\partial} {\tilde{H}} }{ {\partial} \Theta } \right), \\
J_{\Theta}
&=&
- \frac{h}{2} \frac{1}{R} \frac{ {\partial} P }{ {\partial} \Theta }
- \frac{ P }{ \left( \frac{ S_{0} {\lambda} }{ 2 } {\sigma} \right)^{2} + {\mu}^{2} }
 \left( \frac{ \mu }{ R } \frac{ {\partial} {\tilde{H}} }{ {\partial} \Theta }
  + \frac{ S_{0} {\lambda} }{ 2 } {\sigma} \frac{ {\partial} {\tilde{H}} }{ {\partial} R } \right).
\end{eqnarray}
Therefore, the FP equation can be rewritten as follows:
\begin{eqnarray}
  \frac{ {\partial} P }{ {\partial} z }
  = - \frac{1}{R} \left[ \frac{ {\partial} }{ {\partial} R } \left( R J_{R} \right) + \left( \frac{ {\partial} J_{\Theta} }{ {\partial} \Theta } \right) \right] .
\end{eqnarray}

\section{Analysis and the result}
\label{sec:analysis}

We shall now analyze the FP equation.
For this purpose it is sufficient to restrict the
argument for the case that the Hamiltonian depends on $ R $ only:
\begin{eqnarray}
  {\tilde{H}} = H (R) = H .
\end{eqnarray}
Here we assume that function $ P $ does not depend on $ \Theta $,
so the FP equation is written as
\begin{eqnarray}
  \frac{ {\partial} P }{ {\partial} z }
  =
  \frac{ 1 }{ R } \frac{ {\partial} }{ {\partial} R }
    \left[ R \left( \frac{h}{2} \frac{ {\partial} P }{ {\partial} R }
    + \frac{ \mu }{ \left( \frac{ S_{0} {\lambda} }{ 2 } \sigma \right)^{2} + {\mu}^{2} } P \frac{ {\partial} H }{ {\partial} R } \right) \right].
\end{eqnarray}
Then, putting $ P $ as $ P = e^{ - {\Lambda} z } e^{ - \beta H } g $ with
$ \beta = \frac{ 2 \mu }{ h \left[ \left( \frac{ S_{0} {\lambda} }{ 2 } \sigma \right)^{2} + {\mu}^{2} \right] } $,
we arrive at an eigenvalue problem:
\begin{eqnarray}
  \frac{ d }{ dR } \left[ R e^{ - \beta H } \frac{ dg }{ dR } \right]
  =
  - \frac{ 2 \Lambda }{ h } R e^{ - \beta H } g.
\end{eqnarray}
By multiplying $ g $ to both side of this equation, and integrating over $ R $,
we obtain
\begin{eqnarray}
  \int_{0}^{\infty} \left( \frac{ dg }{ dR } \right)^{2} e^{ - \beta H } R \, dR
  =
  \frac{ 2 \Lambda }{ h } \int_{0}^{\infty} g^{2} e^{ - \beta H } R \, dR.
\end{eqnarray}
We see that the eigenvalue $ {\Lambda} = 0 $ requires $ \frac{ dg }{ dR } = 0 $, namely, $ g $ is constant,
which corresponds to the equilibrium state.
To evaluate the eigenvalues, we adopt the variational procedure
such that the following integral is minimized \cite{Brown}:
\begin{eqnarray}
  I = \int_{0}^{\infty} \left( \frac{ dg }{ dR } \right)^{2} e^{ - \beta H } R \, dR,
\end{eqnarray}
under the normalization condition:
\begin{eqnarray}
  \label{eq:normal}
  N = \int_{0}^{\infty} g^{2} e^{ - \beta H } R \, dR = 1.
\end{eqnarray}
Now we look for the {\it lowest excited state},
because the first excited state is dominant in the situation under consideration.
Thus, it is enough to restrict to this state, which is given by a trial form:
\begin{eqnarray}
  g_{1} (R) = A + BR + CR^{2},
\end{eqnarray}
for which we obtain the following expression:
\begin{eqnarray}
  I &=& \int_{0}^{\infty} \left( B + 2CR \right)^{2} e^{ - \beta H } R \, dR  \nonumber \\
    &=& B^{2} J_{1} + 4BC J_{2} + 4C^{2} J_{3} \\
    &\equiv& {\Lambda}_{1} . \nonumber
\end{eqnarray}
Therefore our main problem is to obtain $ {\Lambda}_{1} $.
To carry this out, we use the parabolic form Hamiltonian  $ \tilde{H}_{\rm{F}} = k R^{2} $, which is
inspired from the Faraday effect as given by Eq.~(\ref{eq:hf}).

Let us introduce the parameter $ \alpha = \beta k $
and write $ J_{m} $ as $ J_{m} = \int_{0}^{\infty} R^{m} e^{ - \alpha R^{2} } dR $.
On the other hand, from Eq.~(\ref{eq:normal}), $ N $ becomes
\begin{eqnarray}
  \label{eq:n_condition}
  N = \int_{0}^{\infty} \left( A + BR + CR^{2} \right)^{2} e^{ - \beta H } R \, dR
    = 1.
\end{eqnarray}
We set the orthogonal condition as an additional condition:
$ \int_{0}^{\infty} g_{m} (R) g_{n} (R) e^{ - \beta H } R dR = {\delta}_{m,n} $.
It is possible to assume $ g_{0} (R) = 1 $, hence we obtain
\begin{eqnarray}
  \label{eq:otho_g}
  \int_{0}^{\infty} g_{0} (R) g_{1} (R) e^{ - \beta H } R \, dR
  =
  A J_{1} + B J_{2} + C J_{3}
  =
  0 .
\end{eqnarray}
By eliminating $ A $ from Eqs.~(\ref{eq:n_condition}) and (\ref{eq:otho_g}),
we obtain the equation for $ B $ and $ C $, which is written as $ G(B,C) = 0 $.
In order to obtain $ {\Lambda}_{1} $, we employ the Lagrange multiplier method.
Thus, we calculate the following equations:
\begin{eqnarray}
  \label{eq:lagrange}
  \frac{ {\partial} }{ {\partial} {\chi}_{i} } ( I - \varepsilon G ) = 0
  \ \ \ \
  ( {\chi}_{i} = B, C, \varepsilon ) ,
\end{eqnarray}
where $ \varepsilon $ is a Lagrange multiplier.
By solving Eq.~(\ref{eq:lagrange}), the solution is derived as follows:
\begin{eqnarray}
  ( B , C , {\varepsilon} )
  &=&
  \left( 0 , \pm \sqrt{2} {\alpha}^{3/2} , 4 {\alpha} \right) \nonumber \\
  &=&
  \left( \mp 4 \sqrt{ \frac{ 2 }{ 16 - 5 {\pi} } } {\alpha} , \pm \sqrt{ \frac{ 2 {\pi} }{ 16 - 5 {\pi} } } {\alpha}^{3/2} , \frac{ 4 ( -4 + {\pi} ) }{ -16 + 5 {\pi} } {\alpha} \right).
\end{eqnarray}
Thus we obtain the candidates for the eigenvalue
\begin{eqnarray}
  {\Lambda}_{1} = 4 {\alpha} ,~~~ \frac{ 4 ( -4 + {\pi} ) }{ -16 + 5 {\pi} } {\alpha}.
\end{eqnarray}
Following the variational principle, we should choose the minimum of the above two values,
hence we can settle $ {\Lambda}_{1} = 4 {\alpha} $.
Having determined $ \Lambda_{1} $, we can construct the distribution function $ P(R,z) $ as
\begin{eqnarray}
  P (R,z)
  =
  \left\{ C_{0} g_{0} (R) + C_{1} \exp \left[ - 4 {\alpha} z \right] g_{1} (R) \right\}
  \exp \left[ - {\alpha} R^{2} \right].
\end{eqnarray}
As is seen from this expression, it is enough to take account of the ``ground"  and  the ``first excited state"
only, because the ``higher" order terms are rapidly decreasing.

The diffusion constant is thus given by $ 1 / {\Lambda}_{1} = 1 / 4 {\alpha} $.
By reminding the definition of
$ \alpha = \frac{ 2 \pi \mu {\gamma} }{ ah \left[ \left( \frac{ S_{0} {\lambda} }{ 2 } \sigma \right)^{2} + {\mu}^{2} \right] } $,
we see that the random behavior of the optical spin vortex
may be summarized in the following parameters:
stochastic parameter $ h $, $ {\mu} $, strength of the light $ S_{0} $, wave length $ \lambda $,
topological charge $ \sigma $, size of the vortex $ a $, and the Verde constant $ \gamma $.

It may be significant to make comparison with Eq.~(\ref{eq:sol_rdot}).
On the one hand, Eq.~(\ref{eq:sol_rdot}) describes the classical orbit of the vortex,
whereas $ \alpha $ includes a stochastic effect.
This feature is a sort of the ``classical-quantum" correspondence,
although there is the essential difference between them.
The parameters $ S_{0} $ and $ \lambda $ can be controlled by the external conditions,
whereas the size of vortex $ a $ is a parameter assumed by numerical simulation.

In this viewpoint,
the case for the parabolic form can be solved by an analytical method, which enables us to investigate
the details of the stochastic behavior of the optical spin vortex.
Although we have merely assumed and solved the case of the parabolic form here,
we expect that the consequences from our analyses imply a prospect for the other
cases that are described by more complicated form of the effective Hamiltonian, e.g., the one
caused by the pinning potential.

For the treatment of such general pinning potentials we need to check the accuracy of the variational procedure.
This problem will be left for future study.

\section{Concluding remarks}
\label{sec:conclusion}

We investigated the stochastic aspect of the spin vortex in polarization optics.
Our study is motivated by the viewpoint that the vortex is a significant in wave physics,
because it forms a stable and persistent object during the propagation in spacetime.
This stability may suffer from a presence of the random nature inherent in the optical medium.
The present attempt is based on this general background for the stochastic aspect in wave physics.

We have succeeded in constructing a concise theory of the Langevin and
Fokker-Planck (FP) equations for the evolution of the optical spin vortex with respect to the propagation distance.
The solution of the FP equation provides a simple analytic solution to the diffusive
behavior of the vortex for the specific birefringence that is inspired by the Faraday effect.
The diffusion distance is given in terms of the parameters
that characterize the Faraday effect.
This would provide a clue for the data, which are expected to
be brought about by prospected experiments in polarization optics.

Before concluding, we make the following remark:
The FP equation is essentially non-Hermitian in contrast to the conventional Schr{\"o}dinger equation.
To explore this peculiar nature will be left for future study.

\begin{figure}[t]
  \begin{center}
    \includegraphics[width=\columnwidth]{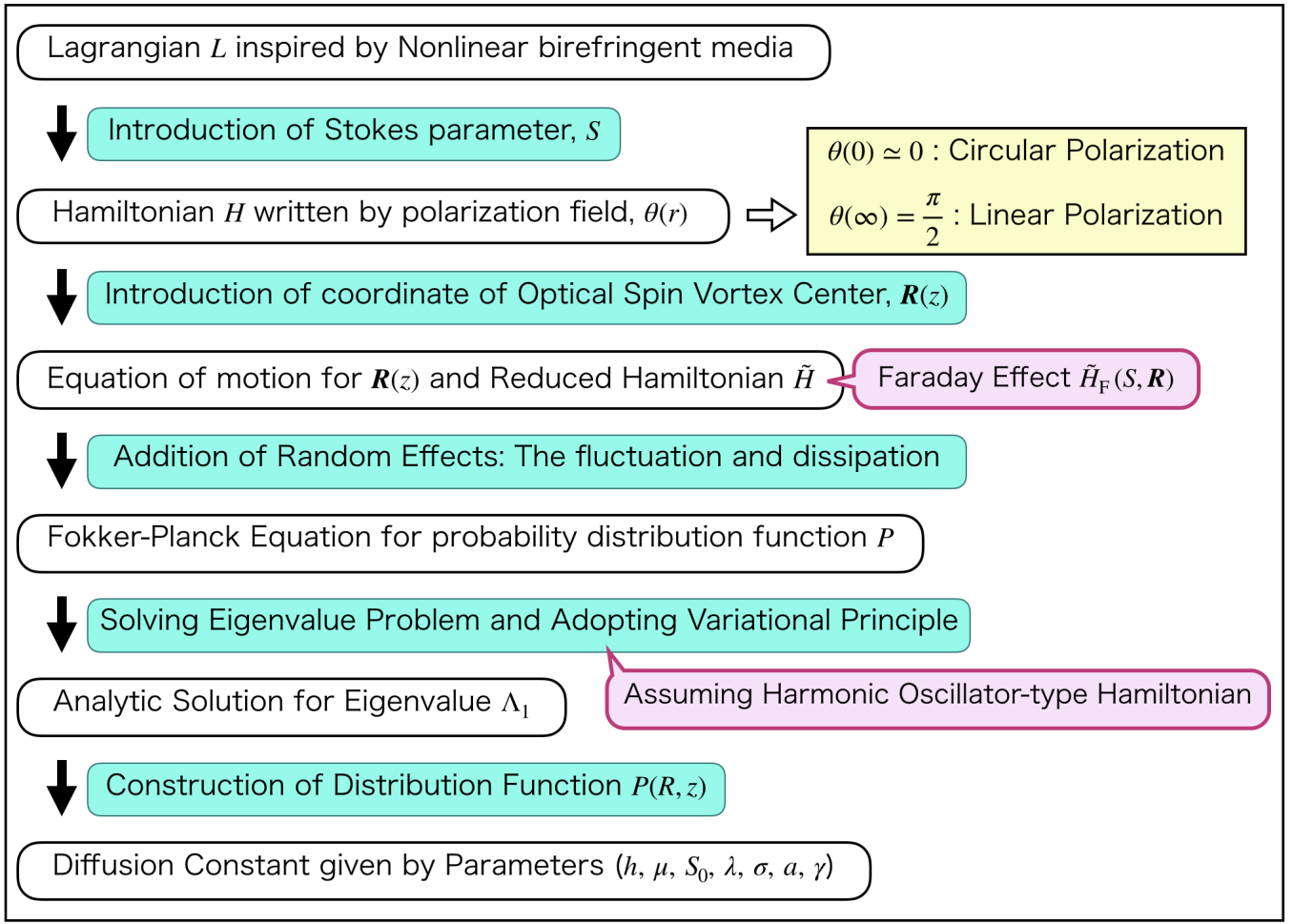}
    \caption{Schematic overview of theoretical development in this paper.}
    \label{fig:sd}
  \end{center}
\end{figure}

Finally, for a convenience for the general readers,  an schematic overview is presented to indicate
the connection between the basic formulations used in the theoretical development:
See, Fig.~\ref{fig:sd}.

\acknowledgments

The work of S. Tsuchida is supported in part by Grant-in-Aid for JSPS Research Fellows, No. 20J00978.

\begin{appendix}

\section{The equation of motion for the vortex center}

We use the relation
\begin{eqnarray}
  \frac{ d }{ dz } \left( \frac{ {\partial} {\tilde{L}} }{ {\partial} \dot{ X } } \right)
  &=&
  \frac{ S_{0} \lambda }{ 2 } \int \left[ \dot{ X } \left( - \frac{ {\partial} v_{x} }{ {\partial} x } \right) + \dot{ Y } \left( - \frac{ {\partial} v_{x} }{ {\partial} y } \right) \right] d^2 x,
  \nonumber \\
  \frac{ {\partial} {\tilde{L}} }{ {\partial} X }
  &=&
  \frac{ S_{0} \lambda }{ 2 }
  \int \left( - \frac{ {\partial} v_{x} }{ {\partial} x } \dot{X} - \frac{ {\partial} v_{y} }{ {\partial} x } \dot{Y} \right) d^2 x
  - \frac{ {\partial} {\tilde{H}} }{ {\partial} X },
 \nonumber
\end{eqnarray}
and hence we obtain the Euler-Lagrange equation for the $ X $ component
\begin{eqnarray}
  \frac{ d }{ dz } \left( \frac{ {\partial} {\tilde{L}} }{ {\partial} \dot{ X } } \right)
    - \frac{ {\partial} {\tilde{L}} }{ {\partial} X }
  =
  \frac{ S_{0} \lambda }{ 2 } \sigma \dot{ Y }  + \frac{ {\partial} {\tilde{H}} }{ {\partial} X }
  =
  0 .
\end{eqnarray}
In addition, the $ Y $ component is obtained in the same form:
\begin{eqnarray}
  \frac{ d }{ dz } \left( \frac{ {\partial} {\tilde{L}} }{ {\partial} \dot{ Y } } \right)
  - \frac{ {\partial} {\tilde{L}} }{ {\partial} Y }
  &=&
  - \frac{ S_{0} \lambda }{ 2 } \sigma \dot{ X } + \frac{ {\partial} {\tilde{H}} }{ {\partial} Y }
  =
  0 .
\end{eqnarray}
Thus, we arrive at Eq.~(\ref{gyration1}).
In the above formula we have introduced the quantity:
$ \sigma = \int_{R^2} ( {\boldsymbol{\nabla}} \times {\boldsymbol{v}} )_{z} \, d^2x $, which represents the ``circulation." Using the expression for the velocity field,
the integrand (which is the {\it vorticity} $ \equiv \omega $) can be written in terms of the angular variables:
$ \omega = {\boldsymbol{\nabla}} \times {\boldsymbol{v}} = \sin \theta ( {\boldsymbol{\nabla}} \theta \times {\boldsymbol{\nabla}} \phi) $
or in terms of the spin field~\cite{Mermin}:
\begin{eqnarray}
  ( {\boldsymbol{\nabla}} \times {\boldsymbol{v}} )_{z}
  =
  {\boldsymbol{l}} \cdot \left( \frac{ \partial {\boldsymbol{l}} }{ \partial x }
  \times \frac{ \partial {\boldsymbol{l}} }{ \partial y } \right),
\end{eqnarray}
where the vector $ {\boldsymbol{l}} \equiv {\boldsymbol{S}} / S_{0} $.
The quantity $ \sigma $ is an optical counterpart of
a topological invariant, which is rewritten as
$ \sigma = \int_{S} \sin \theta \, d \theta \wedge d \phi $,
where $ S $ stands for the area in the Poicar{\'e} sphere.

\section{Derivation of the FP equation}

Here we give the derivation of the FP equation from the functional integral.
Using the Gaussian distribution (\ref{eq:GW}), the transition probability
from the initial vortex center $ {\boldsymbol{R}} (0) $ to the final one $ {\boldsymbol{R}} (z) $
is given by functional integral
\begin{equation}
  K [ {\boldsymbol{R}}(z) \vert {\boldsymbol{R}} (0) ]
  =
  \int_{ {\boldsymbol{R}} (0) }^{ {\boldsymbol{R}} (z) }
  \exp \left[ - \int^{z}_{0} \frac{ {\boldsymbol{\eta}}^{2} (z) }{ 2h } dz \right]
  \mathcal{D} [ {\boldsymbol{\eta}}(z) ] .
\end{equation}
In order to obtain the path integral for the vortex center $ {\boldsymbol{R}} $, we adopt the following steps:
To incorporate the Langevin equation, we insert the expression of the $ \delta $-functional integral
$ \int \delta [ {\boldsymbol{f}} (z) - {\boldsymbol{\eta}} (z) ] \mathcal{D} {\boldsymbol{f}} (z) = 1 $
with $ {\boldsymbol{f}} = \frac{ d {\boldsymbol{R}} }{dz} + {\boldsymbol{A}} $.
After integrating over $ {\boldsymbol{\eta}} $ and $ \rho $ and using the relation
$ \delta[ g(x) ] = \int \exp[ i \rho g(x) ] d \rho $,
one gets
\begin{equation}
  K \sim \int \exp \left[ - \frac{1}{2h} \int_{0}^{z} {\boldsymbol{f}}^{2} (z) \, dz \right] \mathcal{D} {\boldsymbol{f}} .
\end{equation}
Furthermore using the imaginary time $ {\tau} = - i z $~\cite{TK1,TK2},
the propagator is written in the {\it quantum mechanical} path integral form:
\begin{equation}
  \tilde K[ {\boldsymbol{R}} (\tau) \vert {\boldsymbol{R}}(0) ]
  =
  \int \exp \left[ \frac{i}{h} \int \left\{ \frac{1}{2} \left( \frac{ d{\boldsymbol{R}} }{ d \tau } \right)^{2}
  + i {\boldsymbol{A}} \cdot \frac{ d {\boldsymbol{R}} }{d \tau } - W \right\} d \tau \right]
  \mathcal{D} [ {\boldsymbol{R}} ]
\end{equation}
with the potential function $ W =  \frac{ {\boldsymbol{A}}^{2} }{ 2 } -  M h $.
Here the second term comes from the Jacobian,
$ J = {\rm det} \left( \frac{ \delta {\boldsymbol{f}} }{ \delta {\boldsymbol{R}} } \right) $
written in an imaginary time form:  $ J = \exp \left[ \frac{i}{h} \int_{0}^{\tau} M h d \tau \right] $,
where $ M $ is calculated to be
$ M = \frac{1}{2} \frac{ \partial }{ \partial {\boldsymbol{R}} } \cdot {\boldsymbol{A}} $~\cite{TK1}.

The FP equation is thus derived directly by using the above path integral; namely,
by introducing  the ``wave function'' $ \Psi( {\boldsymbol{R}}, \tau) $, we have the integral equation:
\begin{equation}
  \Psi({\boldsymbol{R}}, \tau) = \int K[{\boldsymbol{R}} (\tau) \vert {\boldsymbol{R}} (0) ]
  \Psi({\boldsymbol{R}}, 0 ) d{\boldsymbol{R}}(0) .
\end{equation}
Following the standard procedure of a Feynman path integral, we obtain the Schr{\"o}dinger-type equation:
\begin{equation}
  i h \frac{ {\partial} {\Psi}}{ {\partial} {\tau} }
  =
  \frac{1}{2} \left( {\boldsymbol{p}} - i {\boldsymbol{A}} \right)^{2} \Psi + W \Psi ,
  \label{eq:schrotypeeq}
\end{equation}
where $ {\boldsymbol{p}} = - i h \frac{ {\partial} }{ {\partial} {\boldsymbol{R}} } $.
Now we replace the imaginary time $ {\tau} $ with the original real coordinate $ z $
to get the ``wave function'' $ {\Psi} $ back to the original probability distribution $ P $.
Finally, by using the relation
$ ( {\boldsymbol{\nabla}} \cdot {\boldsymbol{A}} ) P + {\boldsymbol{A}} \cdot {\boldsymbol{\nabla}} P = {\boldsymbol{\nabla}} \cdot( {\boldsymbol{A}} P ) $,
we obtain the FP equation (\ref{eq:FPeq}).

\end{appendix}


\begin{thebibliography}{13}
\bibitem{foot1}
It is to be noted that the reduced form of the fluctuation $ {\boldsymbol{\eta}} $ has a connection with
the original fluctuation $ {\boldsymbol{\xi}} $ by the relation (\ref{eta}), which forms the orthogonal transformation.
Hence the Gaussian white noise holds for $ {\boldsymbol{\eta}} $.
\end{thebibliography}
\end{document}